\documentclass[11pt]{article}
\usepackage{latexsym}

\def\AFOUR{%
\setlength{\textheight}{8.5in}%
\setlength{\textwidth}{5.75in}%
\setlength{\topmargin}{-0.375in}%
\hoffset=-.5in%
\renewcommand{\baselinestretch}{1.17}%
\setlength{\parskip}{6pt plus 2pt}%
}

\AFOUR                                           

\expandafter\ifx\csname amssym.def\endcsname\relax \else\endinput\fi
\expandafter\edef\csname amssym.def\endcsname{%
       \catcode`\noexpand\@=\the\catcode`\@\space}
\catcode`\@=11
\def\undefine#1{\let#1\undefined}
\def\newsymbol#1#2#3#4#5{\let\next@\relax
 \ifnum#2=\@ne\let\next@\msafam@\else
 \ifnum#2=\tw@\let\next@\msbfam@\fi\fi
 \mathchardef#1="#3\next@#4#5}
\def\mathhexbox@#1#2#3{\relax
 \ifmmode\mathpalette{}{\m@th\mathchar"#1#2#3}%
 \else\leavevmode\hbox{$\m@th\mathchar"#1#2#3$}\fi}
\def\hexnumber@#1{\ifcase#1 0\or 1\or 2\or 3\or 4\or 5\or 6\or 7\or 8\or
 9\or A\or B\or C\or D\or E\or F\fi}

\font\tenmsa=msam10
\font\sevenmsa=msam7
\font\fivemsa=msam5
\newfam\msafam
\textfont\msafam=\tenmsa
\scriptfont\msafam=\sevenmsa
\scriptscriptfont\msafam=\fivemsa
\edef\msafam@{\hexnumber@\msafam}
\mathchardef\dabar@"0\msafam@39
\def\dashrightarrow{\mathrel{\dabar@\dabar@\mathchar"0\msafam@4B}}
\def\dashleftarrow{\mathrel{\mathchar"0\msafam@4C\dabar@\dabar@}}

\def\ulcorner{\delimiter"4\msafam@70\msafam@70 }
\def\urcorner{\delimiter"5\msafam@71\msafam@71 }
\def\llcorner{\delimiter"4\msafam@78\msafam@78 }
\def\lrcorner{\delimiter"5\msafam@79\msafam@79 }
\def\yen{{\mathhexbox@\msafam@55}}
\def\checkmark{{\mathhexbox@\msafam@58}}
\def\circledR{{\mathhexbox@\msafam@72}}
\def\maltese{{\mathhexbox@\msafam@7A}}
\def\circledS{{\mathhexbox@\msafam@73}}

\font\tenmsb=msbm10
\font\sevenmsb=msbm7
\font\fivemsb=msbm5
\newfam\msbfam
\textfont\msbfam=\tenmsb
\scriptfont\msbfam=\sevenmsb
\scriptscriptfont\msbfam=\fivemsb
\edef\msbfam@{\hexnumber@\msbfam}
\def\Bbb#1{{\fam\msbfam\relax#1}}
\def\widehat#1{\setbox\z@\hbox{$\m@th#1$}%
 \ifdim\wd\z@>\tw@ em\mathaccent"0\msbfam@5B{#1}%
 \else\mathaccent"0362{#1}\fi}
\def\widetilde#1{\setbox\z@\hbox{$\m@th#1$}%
 \ifdim\wd\z@>\tw@ em\mathaccent"0\msbfam@5D{#1}%
 \else\mathaccent"0365{#1}\fi}
\font\teneufm=eufm10
\font\seveneufm=eufm7
\font\fiveeufm=eufm5
\newfam\eufmfam
\textfont\eufmfam=\teneufm
\scriptfont\eufmfam=\seveneufm
\scriptscriptfont\eufmfam=\fiveeufm
\def\frak#1{{\fam\eufmfam\relax#1}}

\csname amssym.def\endcsname

\makeatletter
\def\section{\@startsection {section}{1}{\z@}{-3.5ex plus -1ex minus
 -.2ex}{2.3ex plus .2ex}{\large\sc}}
\def\subsection{\@startsection{subsection}{2}{\z@}{-3.25ex plus -1ex minus
 -.2ex}{1.5ex plus .2ex}{\normalsize\sc}}
\makeatother

\makeatletter
\@addtoreset{equation}{section}

\makeatother

\newcommand{\nc}{\newcommand}
\newcommand{\rnc}{\renewcommand}

\nc{\be}{\begin{equation}}
\nc{\ee}{\end{equation}}
\nc{\bea}{\begin{eqnarray}}
\nc{\eea}{\end{eqnarray}}

\nc{\trac}[2]{{\textstyle\frac{#1}{#2}}}

\nc{\ex}[1]{\mbox{e}^{\,\textstyle#1}}

\nc{\CC}{\Bbb{C}}
\nc{\HH}{\Bbb{H}}
\nc{\PP}{\Bbb{P}}
\nc{\RR}{\Bbb{R}}
\nc{\ZZ}{\Bbb{Z}}
\nc{\II}{\Bbb{I}}
\nc{\EE}{\Bbb{E}}

\rnc{\a}{\alpha}
\rnc{\b}{\beta}
\rnc{\d}{\delta}
\nc{\ga}{\gamma}
\nc{\f}{\phi}
\nc{\e}{\eta}
\rnc{\c}{\chi}
\nc{\eps}{\epsilon}
\nc{\om}{\omega}
\nc{\Om}{\Omega}

\nc{\symx}{\circledS}
\newsymbol\smallsmile 1360
\newsymbol\smallfrown 1361
\nc{\ad}{\mathop{\mbox{ad}}\nolimits}
\nc{\tr}{\mathop{\mbox{tr}}\nolimits}
\nc{\Tr}{\mathop{\mbox{Tr}}\nolimits}
\nc{\Det}{\mathop{\mbox{Det}}\nolimits}
\rnc{\det}{\mathop{\mbox{det}}\nolimits}
\nc{\rk}{\mathop{\mbox{rk}}\nolimits}
\nc{\sign}{\mathop{\mbox{sign}}\nolimits}
\nc{\del}{\partial}
\nc{\diag}{\mbox{diag}}
\nc{\ra}{\rightarrow}
\nc{\Ra}{\Rightarrow}
\nc{\LRa}{\Leftrightarrow}
\nc{\lra}{\leftrightarrow}
\nc{\ot}{\otimes}
\rnc{\ss}{\subset}
\nc{\nul}{\noindent\underline}
\nc{\non}{\nonumber\\}
\nc{\mat}[4]{\left(\begin{array}{cc}#1&#2\\#3&#4\end{array}\right)}
\rnc{\lg}{\frak{g}}

\begin{document}


\def\ci{\cite}
\def\ov{\over}
\def\ha{{ 1\ov 2}}
\def\four{{1 \ov 4}}
\def\td{\tilde}
\def\ff{{\rm f}}
\def\pp{{\rm p}}
\def\qq{{\rm q}}
\def\ww{{\rm w}}
\def\hh{{\rm h}}
\def\kk{{\rm k}}
\def\xx{{\rm x}}
\def\yy{{\rm y}}
\def\zz{{\rm z}}
\def\uu{{U}}
\def\vv{{V}}
\def\uuu{{\rm u}}
\def\vvv{{\rm v}}
\def\thth{\rm \theta}
\def\const{{\rm const}}
\def\ep{\epsilon}
\def\lc{{light-cone}}
\def\bZ{\ZZ}
\def\bR{\RR}
\def\bR{\RR}
\def\bE{\EE}
\def\g{\gamma}
\def\foot{\footnote}
\newcommand{\rf}[1]{(\ref{#1})}
\def\la{\label}
\def \p {\phi}
\def \bi{\bibitem}
\def \s{\sigma}
\def\E{{\cal E}}
\def\H{{\cal H}}
\def\D{{\rm D}}
\def \m {\mu}
\def \aa {{\rm a}}



\rightline{}
\rightline{}

\begin{center}
{\Large\sc Power-law singularities  in  string theory and M-theory}
\end{center}
\vspace{0.3cm}

\begin{center}
{\large G.\ Papadopoulos
}
\end{center}

\vskip 0.1 cm

\centerline{\it ${}$ Department of Mathematics}
\centerline{\it King's College
London}
 \centerline{\it London WC2R 2LS, U.K. }
 \centerline{\small gpapas@mth.kcl.ac.uk}

\vskip 5.0 cm

\begin{center}
{\bf Abstract}
\end{center}
\vskip -0.1 cm

We extend the definition of  the Szekeres-Iyer power-law
singularities  to supergravity, string and
M-theory backgrounds, and find that
are characterized by Kasner type exponents.
The near singularity geometries of brane and some intersecting brane
backgrounds are investigated and  the exponents are computed.
The Penrose limits of some of these power-law singularities have
profiles $A\sim {\rm u}^{-\gamma}$ for $\gamma\geq 2$. We find the range
of the exponents for which $\gamma=2$ and the frequency squares
are bounded by $1/4$.  We propose some qualitative tests for deciding
 whether a null or timelike spacetime singularity can be resolved within string theory
 and M-theory based on the
 near singularity geometry and its Penrose limits.

\newpage
\begin{small}
\tableofcontents
\end{small}

\newpage
\setcounter{footnote}{0}

\section{Introduction}

Szekeres and Iyer \cite{SI,CS}~investigated the near singularity geometries of
four-dimensional spherically symmetric solutions to the Einstein
equations. They mostly focused  on four-dimensional geometries
with singularities of power-law type
 that arise during gravitational
collapse, like the Lemaitre-Tolman-Bondi dust
collapse metrics and Lifshitz-Khalatnikov
singularities. It turns out that
the geometry near the singularities can be described as
\be
ds^2=- 2 x^\pp dudv+ x^\qq d\Omega^2_{2}~,
\la{si}
\ee
where
\be
x=k u+\ell v~,~~~~~~\ell, k=0, \pm1~.
\ee
The Kasner type exponents $\pp,\qq$  characterize the behaviour of
 the geometry near the singularity
at $x=0$.

It has been observed  that  string modes in plane waves
with profiles $A\sim \uuu^{-\gamma}$ for $\gamma<2$ \cite{sanchez, david} and for $\gamma=2$
with frequency squares $\omega^2<1/4$ \cite{prt}
can be extended across the singularity at $\uuu=0$; for similar results in field theory see
\cite{turok,giveon}. This is despite the fact  plane waves with such profiles
 $\gamma>0$ have
a curvature singularity at $\uuu=0$ \cite{horow, marolf}.
An interpretation
of this is that strings have a smooth propagation\footnote{However there
are processes which are singular. For example for particular
states there is an infinite string mode production near the
singularity \cite{horow, prt}.} in such backgrounds.   Motivated  by this,
the authors of \cite{papad1, papad2} computed the Penrose limits of the near
singularity geometries (\ref{si}) and found that $A\sim \uuu^{-\gamma}$, $\gamma\geq2$,
extending the results of \cite{bfp, fuji,ryang, patricot, kunze}.
In addition, they determined the range of the exponents
for the wave profiles to behave as  $A\sim \uuu^{-2}$ with $\omega^2<1/4$.
It was found that all
geometries that satisfy (but not saturate) the dominant energy condition
have Penrose limits with such profiles.

In this paper, we generalize the Szekeres-Iyer
power-law singularities for string theory and M-theory backgrounds.
This generalization
applies to backgrounds  with metrics and n-form
field strengths which locally can be written as
\bea
ds^2&= &g_{mn}(y) dy^m dy^n+ \sum_{i} G^2_i(y) ds^2_{(i)}
\cr
F_n&=& {\rm F}_{mn}(y) dy^m\wedge dy^n\wedge  \omega
+ {\rm F}_m(y) dy^m\wedge \chi
+{\rm F}(y)\tau~,
\la{gmet}
\eea
where $g$ is a two-dimensional Lorentz metric $m,n=1,2$, and
$ds^2_{(i)}$ are smooth metrics and $\omega, \chi, \tau$
are forms on the Riemannian manifolds ${\cal M}_i$
{\sl independent} from the coordinates $y^m$. In addition, we assume that the
spacetime has singularities of {\sl codimension one}, ie every singularity is
specified by a single function $C=C(y)$.
Demanding that the singularities are of power-law type,
the near singularity geometry is
characterized by Kasner type exponents both for the metric
and the form-field strengths. The singularities that arise
are timelike, spacelike or null.

A study of  generic spacelike singularities in ten- and
eleven-dimensional supergravities, following earlier work in general
relativity \cite{bkl},
has revealed that their behaviour is chaotic \cite{damour};
 for a review see \cite{nicolai}. This chaotic behaviour
resembles that of a mixmaster universe where the Kasner
exponents change infinite many times along  independent spatial
directions.
Because of this our results for spacelike singularities
are not generic. A similar analysis for weak null singularities
in general relativity has been done in \cite{ori1}. Nevertheless,
 the method we propose works
well for some special backgrounds with spacelike and null singularities
and also is applicable to backgrounds with timelike  singularities.

We demonstrate that some brane backgrounds have singularities
of power-law type and give their near singularity geometries. In particular,
we compute the exponents of the near singularity geometries of
the fundamental string and Dp-brane, $p\not=3$, $p\leq 6$, backgrounds.
It turns out that the singularities of fundamental string and Dp-branes, $p\leq5$,
are null while the singularity of the D6-brane is timelike.
The metric of NS5-brane in the
string frame, the D3-brane and  M5-brane  are not singular\cite{gibbons}.

We also analyze the Penrose limits
of the near singularity geometries of all
 power-law singularities. Some power-law singularities have
 diagonal plane wave profiles $A$
which  behave as $A\sim \uuu^{-\gamma}$, $\gamma\geq2$, where
$\uuu$ is the affine parameter of a null geodesic
and the singularity of the original spacetime is located at $\uuu=0$.
We find the conditions on the exponents of these near singularity geometries
for the wave profile to have behaviour  $\gamma=2$ and the frequency squares
to be bounded by $\omega^2\leq 1/4$. There are power-law
singularities for which the Penrose
limits have   non-diagonal
plane wave profiles. For these we give the
 plane wave metric in Rosen coordinates.
Some of these Penrose limits may lead to homogeneous plane 
waves with rotation \cite{mmol}.

We  propose a number of qualitative tests to decide whether
 a spacetime singularity
can be resolved in string theory and M-theory. These tests mainly rely (i)
on whether
the near singularity geometry of a background can be identified with that of
another singularity which has a well known description within string theory, (ii)
 on the assumption that the Penrose limits of a background can be taken
 in a regular way, and (iii)
 on whether string theory and M-theory is singular
 or well-defined at the Penrose limits.
 If a string   background is singular but the singularity has
 a well-known interpretation, eg it has the near singularity geometry
 of a brane, and string theory  is well-defined at all its Penrose
 limits, then these can serve as an indication
 that this singularity can
 be resolved within string theory.  Though other
 tests should also be performed before it is decided whether
 string theory  is well-defined in such background,
 see eg \cite{Liu, giveonb, costa}.  Alternatively,
 if  the near singularity geometry  of a background  does not have
 an interpretation within string theory and string theory
 is singular at a Penrose limit, eg string modes cannot propagate through the
  Penrose limit singularity, then we argue that such background
 is singular.

As an application of the above tests, we consider backgrounds
  with diagonal wave profiles $A\sim \uuu^{-\gamma}$.
We categorize such singularities  into
three types mild, marginal and severe.
{\it Mild} singularities are those for which all the Penrose limits near the
singularity have plane wave profiles $A\sim \uuu^{-\gamma}$ with $0<\gamma<2$.
{\it Marginal} singularities are those for which all the Penrose limits near the
singularity have plane wave profiles $A\sim \uuu^{-2}$.
{\it Severe} singularities are those for which all the Penrose limits near the
singularity have plane wave profiles $A\sim \uuu^{-\gamma}$ with $\gamma> 2$.
We shall argue that
backgrounds which have timelike and null singularities of  the marginal type with
frequency squares $\omega^2>1/4$ and singularities of severe type may be
singular in string theory. This is based on our result that the near singularity
geometries of branes 
are either timelike or null and their Penrose limits have marginal
singularities with frequency squares
$\omega^2\leq 1/4$ \cite{bfp, fuji, ryang}. It is also known that  string modes
propagate  across the mild \cite{sanchez} and marginal
singularities \cite{prt} of plane
waves with frequency squares $\omega^2<1/4$
but they are singular for the rest of planes waves with the
 above profiles \cite{sanchez}.
It is also likely that these results generalize to spacelike singularities.

This paper is organized as follows: In section two, the definition
of the near singularity
geometries is given for (singular) string and M-theory
backgrounds and describe how the near singularity geometry is characterized
by exponents.  In section three, we find the near singularity geometries of infinite
planar brane solutions of ten- and eleven-dimensional supergravities. We also
give the near singularity geometries of some intersecting brane configurations.
In section four, we explain how the near singularity geometries of null
and timelike power-law singularities and
their Penrose limits can be used to provide some qualitative tests
on whether string theory and M-theory is singular in certain backgrounds.
In appendix A, we compute the Penrose limits of power-law singularities that
arise in string and M-theory. We argue that some of them are associated with
homogeneous singular plane waves with rotation. In appendix B, we describe
the various Penrose limits that can be used for backgrounds with $\alpha'$
corrections.

\section{Codimension one power-law singularities in string theory and M-theory}

\subsection{Near singularity metrics}

 We shall extend the Szekeres-Iyer definition of near singularity
 geometries for power-law
 singularities to the class
 of string and M-theory backgrounds for which the the metric can be written as
\be
ds^2= \gamma_{mn}(y) dy^m dy^n+ \sum_{i} G^2_i(y) ds^2_{(i)}~, ~~~~~m,n=0,1~.
\ee
In addition, we assume that some of the components of $\gamma$ and $G^2_i$ vanish
or are infinite at
\be
C(y)=0
\ee
 At such a hypersurface,
the metric $ds^2$ can be singular. Since the singularity is specified by a
single equation, it is of codimension one.

It is well-known that all two-dimensional metrics are conformally flat. Choosing
the conformal gauge, we can write the metric $ds^2$  as
\be
ds^2=- 2 K^2({\uu},{\vv})  d\uu d\vv+ \sum_{i} G^2_i({\uu},{\vv}) ds^2_{(i)}~.
\la{cgauge}
\ee
The transformations that preserve this form of the metric are two-dimensional
conformal transformations in the coordinates $(\uu, \vv)$ and
 the diffeomorphims of ${\cal M}_i$.

The equation for the singularity can now be written as
\be
C({\uu},{\vv})=0~.
\la{seq}
\ee
Suppose that in some conformal coordinates $(\uu, \vv)$ the equation for the
singularity can be written as
\be
C({\uu},{\vv})= m(\uu,\vv) \biggr(k  f({\uu})+ \ell h({\vv})\biggl)^\alpha=0~,~~~~
~~k, \ell=0, \pm 1~, ~~~~~\alpha\in \bR~,
\ee
where $m$ is a regular function and $m\not=0, \infty$ for $k f(\uu)+\ell  h(\vv)=0$.
If the singularity equation can be written as above\footnote{This is not a strong assumption.
For most singularities
$C=0$ can be solved using the inverse function theorem as $\uu=c(\vv)$ and then use
a conformal transformation in $\vv$ to find the coordinate $x$. However the assumption
that the singularity equation is of the above form an  advantage which enable us
to describe  singularities that are located at infinity.}, then there is always a coordinate
$x$ such that $x>0$ and the singularity is located at $x=0$. To show this,
there are several cases to consider the following:

 (i)~~~ If $\eta=k \ell\not=0$ and $\alpha>0$,
 we perform the conformal transformation $u=f(\uu)$, $v=h(\vv)$ and define $x=k\uu+\ell \vv$.
 In this new
coordinate, the
singularity is located at $x=0$. Moreover we choose $k,\ell$ such that
 $x>0$.

(ii)~~ If $\eta=k \ell\not=0$ and $\alpha<0$. We first define conformal coordinates
$\uu'=f(\uu), \vv'=h(\vv)$. It is now clear that the singularity lies at infinity, ie
either at $\uu'=\pm\infty$ and/or at $\vv'=\pm \infty$.
In the latter case, we perform the conformal transformation $u=\uu'$ and $v=1/\vv'$. The
equation for the singularity can be rewritten as
\be
C(u,v)= m(u, {1\over v}) \biggr({\ell v\over \eta uv+1}\biggl)^{-\alpha}~.
\ee
We define a new coordinate $x=\ell v$ and choose $\ell$ such that $x>0$.
The singularity is  located
at $x=0$. The case that the singularity lies at $\uu'=\pm \infty$ can be treated
in a similar way. 

(iii)~ If $\eta=0$, because say  $k=0$, and $\alpha>0$, we again define  conformal
coordinates $v=h(\vv)$ and $u=\uu$ and
 $x=\ell v$. Again $x>0$ with an appropriate choice of $\ell$ and
  the singularity is located at $x=0$. The case with
$k\not=0$, $\ell=0$ can be similarly treated.

(iv)~~ if $\eta=0$, say because $k=0$, and $\alpha<0$, we again define
conformal coordinates $\uu'=\uu$, $\vv'=h(\vv)$. The singularity lies at infinity,
$\vv'=\pm \infty$. In this case, we perform another conformal transformation
$u=\uu'$, $v=1/\vv'$ and define  $x=\ell v$.  Again $x>0$ with
an appropriate choice of $\ell$ and
  the singularity is located at $x=0$. The case with
$k\not=0$, $\ell=0$ can be similarly treated.

We have seen that for all types of singularities singularities we have considered,
there are conformal coordinates $(u, v)$ and a coordinate
\be
x=k u+\ell v~, ~~~~\eta=k \ell=0, \pm1~,~~~~x>0
\ee
which has the property that the singularity equation can be written as
\be
C=\tilde m(u,v) x^\alpha~,~~~~\alpha>0~,
\la{ccss}
\ee
where $\tilde m$ regular function and $\tilde m\not=0, \infty$ at $x=0$.
Thus the singularity is located at $x=0$. Singularities for which $x=k u+\ell v$
with $\eta=k \ell\not=0$ are either spacelike ($\eta=1$) or
 timelike $(\eta=-1)$ while singularities
for which $\eta=0$ are null.

The metric in the $u,v$ conformal coordinates is
\be
ds^2= -2 L^2(u, v)  du dv+ \sum_{i} G^2_i( u, v) ds^2_{(i)}~,
\la{scgauge}
\ee
where the new conformal factor $L^2$ can be easily computed from $K^2$ and the
conformal transformation necessary to bring the equation for the singularity
in the form (\ref{ccss}).
The components of the metric $L^2$, $G_i^2$ can be expressed as functions of the
new coordinate  $x$ and either $u$, if $k=0$ and $k, \ell\not=0$,  or $v$, if $\ell=0$.
Without loss of generality, we assume that $L^2$, $G_i^2$ can be expressed
as functions of $x, u$. The analysis for the other case is similar.
Since we are concerned with the behaviour of the metric near the singularity $x=0$,
we expand $L^2$, $G_i^2$ in power-series in $x$.
We {\it demand } the singularity at $x=0$ is of  power-law
type, ie  $L^2, G_i^2$ have an expansion in $x$ of the form
\bea
L^2&=& l(u) x^{\pp}+  l_1(u) x^{\pp_1}+\dots 
\cr
G_i^2&=& g_i(u) x^{\ww_i}+ g_{i1}(u) x^{\ww_{i1}}+\dots
\eea
for some regular functions $\{l, l_1, \dots\}$
and $\{g_i, g_{i1}, \dots\}$  and $l, g_i>0$,
where
$\pp<\pp_1<\dots$
and
$\ww_i<\ww_{i1}<\dots$.
Writing $L^2= e^A$ and $G_i^2= e^{B_i}$, we have that
\bea
A&=&\pp \log x+ \alpha(u, x)
\cr
B_i&=& \ww_i \log x+\beta_i(u, x)
\eea
where $\alpha$, $\beta_i$ are regular functions as $x\rightarrow 0$.
The `near singularity geometry' of the metric (\ref{scgauge}) is {\it defined}
as
\be
d\bar s^2=- 2 x^\pp dudv+  \sum _i x^{\ww_i} ds^2_{(i)}~.
\la{pl}
\ee
The Kasner type exponents $\pp, \ww_i$  characterize
the power-law   singularity. Observe that although we have started from
a special class of metrics, the metric (\ref{pl}) is the most general
power-law metric for spacelike, timelike and null singularities.

If one takes as the equation for the singularity $x=0$, then there is no residual
diffeomorphism invariance in the $y^m$ coordinates that leaves both
the form of the metric (\ref{pl}) and the equation of
singularity invariant.
However, there are residual transformations that only leave the form of the metric
invariant. As a  result, there is a relation
between the Kasner type exponents $\pp, \ww_i$. We shall not pursue this further
because the form of the metric (\ref{pl}) is sufficient for our purpose.
Note that in the null case, the sign of the $u,v$ component of the metric can change
with a coordinate transformation so it has not an invariant meaning.

If $\eta\not=0$, we define a new  coordinate $y=k u-\ell v$ and rewrite the metric as
\be
ds^2=-{1\over2} \eta x^\pp (dx^2-dy^2)+  \sum _i x^{\ww_i} ds^2_{(i)}~,
~~~~\eta=k \ell~.
\la{meta}
\ee
It is now clear that the singularities with $\eta=1$ are spacelike while those
with $\eta=-1$ are timelike. As we have mentioned (\ref{meta}) does not describe
the generic behaviour of spacelike singularities in supergravity theories because
they are not of power-law type but exhibit mixmaster or chaotic behaviour \cite{damour, nicolai}.
 A similar analysis has been done for weak
null singularities in \cite{ori1} and some applications have been found
in \cite{ori2, bdm}.
However (\ref{meta}) is general enough
to describe all singularities of supergravity theories which are
of power-law type.

In many examples that we shall investigate later, the metric (\ref{scgauge}) is of special
form. This in turn  leads to a near singularity metric which can be written as
\be
d\bar s^2= 2 x^\pp dudv+ \sum_{i} x^{\ww_i} ds^2(\bR^{n_i})+ x^\qq d\Omega^2_m~.
\la{bgsi}
\ee
The round m-sphere metric $d\Omega_m^2$ appears in many brane configurations as
the sphere of the overall transverse space at infinity.

\subsection{Near singularity field strengths}

M-theory and string theory backgrounds apart from the metric  have other `active'
fields, like a dilaton or a form field strength.
 Therefore singular backgrounds are  those for which either
the metric  or one of the other fields develops a singularity.
To define the power-law singularities in the presence of other fields, we consider
a background which has a non-vanishing  n-form field strength
$F_n$ which is singular at the codimension one singularity
(\ref{seq}). There are several definitions of what a  singularity is in relation
to the metric, for example one can take the definition that the
 spacetime is geodesically incomplete.
In the case of other fields, one can also use various definitions.
Here with singularity of a form field strength, we mean a region in spacetime that the
form diverges. This is not an invariant definition but it will suffice
for our purpose\footnote{An invariant definition of a singularity for $F_n$ is to
take the length of
$F_n$ to diverge.}.
In the conformal coordinates $(u,v)$ defined in the previous
section, the n-form field strength (\ref{gmet}) can be expanded as
\be
F_n= {\rm F}_1(u,v) du\wedge dv\wedge  \omega+ {\rm F}_2(u,v)
 du\wedge \chi+ {\rm F}_3(u,v)dv \wedge \psi+
{\rm F}_4(u,v)\tau~,
\ee
where $\omega, \chi, \psi$ and $\tau$ are forms  of appropriate degree in the
remaining coordinates independent from $u,v$ and $\{{\rm F}_s, s=1, \dots, 4\}$
are functions of $u,v$. (If $F_n$ has degree one, then $\omega=0$ and if $F_n$
has degree zero $\omega=\chi=\psi=0$ and $\tau$ is a scalar.)
{}For power-law singularities,  ${\rm F}_s$ have an expansion
\be
{\rm F}_s=f_s(u) x^{\ff_s}+  f_{1s}(u) x^{\ff_{s_1}}+\dots ,~~~~s=1,\dots,4~,
\ee
where $\ff_s<\ff_{s_1}<\dots$, $f_s(u)\not=0$ . The form field strength of the near
singularity geometry is defined as
\be
\bar F_n=x^{\ff_1} du\wedge dv\wedge  \omega+x^{\ff_2} du\wedge \chi+ x^{\ff_3} dv \wedge \psi
+ x^{\ff_4}\tau~
\la{gff}
\ee
in analogy  with the definition of the near singularity metric.

The definition we have given for the near singularity
geometry for power-law singularities captures the leading
behaviour of the metric and form field strengths
as one approaches the singularity. We have not verified that
this leading order contribution solves the original field equations.
Nevertheless, the above definition suffices for our purpose
to investigate the qualitative properties of the singularities
and their Penrose limits.

\section{Power-law singularities and brane configurations}

\subsection{Dp-branes}

It is well-known that all Dp-brane backgrounds
\cite{DuffLu5Brane, DuffLuD3, DuffLu5Brane, HorowitzStrominger}, apart from the D3-brane, are
singular in the string frame. The  Dp-brane background  can be written as
\bea
ds^2&=& H^{-{1\over2}} ds^2(\bR^{1,p})+ H^{{1\over2}} (dr^2+ r^2 d\Omega^2_{8-p})~,
~~~~~~~~~~H=1+{ Q_p\over r^{7-p}}~, ~~~~~~~~~p\leq 6
\cr
F_{p+2}&=& d{\rm vol}(\bR^{1,p})\wedge dH^{-1}
\cr
e^{2\phi}&=&H^{3-p\over2}~,
\eea
where $F_{p+2}$ is the (p+2)-form field strength, $p\not=3$, and $\phi$ is the dilaton.
The mass per unit volume
of the Dp-brane, $Q_p$, does not contribute in the  exponents of the singularity,
so  without loss of generality we can set $Q_p=1$.
The geometry near the singularity $r=0$ is
\be
ds^2= r^{{7-p\over2}} ds^2(\bR^{1,p})+ r^{-{7-p\over2}}(dr^2+ r^2 d\Omega^2_{8-p})~.
\ee
As we have explained in the previous section to find the metric near the
singularity,
we write the two-dimensional metric
\be
ds^2_{(2)}= -r^{{7-p\over2}} dt^2+  r^{-{7-p\over2}} dr^2
\ee
in the conformal gauge.
{}For this we first change coordinates as $r=w^\alpha$ which give
\be
ds^2_{(2)}=- w^{{7-p\over2}\alpha} dt^2+r^{-{7-p\over2}\alpha+2\alpha-2} \alpha^2 dw^2
~.
\ee
We then  set $\alpha=-{2\over 5-p}$, $p\not=5$, and  the Dp-brane background
can be written as
\bea
ds^2&=& w^{-{7-p\over5-p}}\biggr(- dt^2+ {4\over (5-p)^2} dw^2
+ds^2(\bR^p)\biggl)+ w^{{3-p\over5-p}}
d\Omega^2_{8-p})
\cr
F_{p+2}&=&-{14-2p\over 5-p} w^{-{19-3p\over 5-p}}~ d{\rm vol}(\bR^{1,p})\wedge dw
\cr
e^{2\phi}&=& w^{(3-p) (7-p)\over 5-p}
~.
\eea
 We further
set $z={2\over |p-5|}w$ which gives
\bea
ds^2&=&\biggr({|p-5|\over 2}\biggl)^{-{7-p\over5-p}}\biggr[ z^{-{7-p\over5-p}}
\biggr(-dt^2+ dz^2+ds^2(\bR^p)\biggl)+
\biggr({|p-5|\over 2}\biggl)^2 z^{{3-p\over5-p}}
d\Omega^2_{8-p}\biggl]
\cr
F_{p+2}&=&-\biggr({14-2p\over 5-p}\biggl) \biggr({|p-5|\over 2}\biggl)^{-{14-2p\over 5-p}}
z^{-{19-3p\over 5-p}}~ d{\rm vol}(\bR^{1,p})\wedge dz
\cr
e^{2\phi}&=& \biggr({|p-5|\over 2}\biggl)^{(3-p) (7-p)\over 5-p} z^{(3-p) (7-p)\over 5-p}
~.
\eea
The above metric has singularities at $z=0$ and $z=\infty$.
 For D6-branes, $r={1\over 4}z^2$
and the singularity at $r=0$ is located at $z=0$. 
The singularity which is located at
$z=\infty$ is an artifact of the approximation we made
 by replacing the the harmonic function
$H$ with its near singularity value $1/r$ and so it is not a singularity of the D6-brane.
Therefore, we have that the singularity is at 
$C=r={1\over 4}z^2=0$, which can be
solved by setting $z=0$. Because of this, we set $x=z=u+v$, $t=v-u$,
($u={1\over2}(-t+x)$, $v={1\over 2}(t+x)$) to find that the near singularity
background is
\bea
d\bar s^2&=&2 x dudv+
x ds^2(\bR^6)+ x^3
d\Omega^2_{2}
\cr
\bar F_{p+2}&=& -
x~  du\wedge dv\wedge d{\rm vol}(\bR^{6})
\cr
e^{2\bar\phi}&=& {1\over8} x^{3}
~.
\eea
The  exponents of the power-law
singularities of the D6-brane  are
\bea
ds^2&:&{\rm p}=1~,~~~~~~{\rm w}=1~,~~~~~~{\rm q}=3
\cr
F_8&:& \ff=1
\cr
\phi&:& {\rm d}=3
\eea
The D6-brane singularity is a timelike singularity.

{}For the  Dp-branes, $p<5$, we have that $r=w^{-{2\over 5-p}}= {|p-5|\over 2}^{-{2\over 5-p}}
z^{-{2\over 5-p}}$ and so the singularity at $r=0$ is located at $z=\infty$.
To proceed define the conformal coordinates
 $\uu={1\over2}(-t+z)$, $\vv={1\over 2}(t+z)$.
In terms of these coordinates, the singularity equation $C$ becomes
\be
C={1\over z^{{2\over 5-p}}}={1\over (\uu+\vv)^{{2\over 5-p}}}=0~.
\ee
The solutions to this equation are either $\uu=+\infty$ or $\vv=+\infty$.
The two cases are symmetric and so without loss of generality we take
the singularity to lie at $\vv=+\infty$. The other case whether the
singularity lies at $\uu=+\infty$ leads to a near singularity geometry
with the same exponents.
As we have explained to bring the singularity from infinity to the
 origin, we perform the conformal
transformation $\uu=u$, $\vv=1/ v$. After this transformation,
the singularity equation is
\be
C=\biggr({v\over u v+1}\biggl)^{{2\over 5-p}}=0~.
\ee
and so we choose, $x=v$. The Dp-brane background becomes
\bea
ds^2&=&2 \biggr({|p-5|\over 2}\biggl)^{-{7-p\over5-p}}
\biggr[-2 z^{-{7-p\over5-p}} v^{-2} dudv+
{1\over 2} z^{-{7-p\over5-p}}ds^2(\bR^p)+{1\over 2}
\biggr({p-5\over 2}\biggl)^2 z^{{3-p\over5-p}}
d\Omega^2_{8-p}\biggl]
\cr
F_{p+2}&=& (-1)^{p+1}2
 \biggr({14-2p\over 5-p}\biggl) \biggr({|p-5|\over 2}\biggl)^{-{14-2p\over 5-p}}
z^{-{19-3p\over 5-p}} v^{-2}~  du\wedge dv\wedge d{\rm vol}(\bR^{p})
\cr
e^{2\phi}&=& \biggr({|p-5|\over 2}\biggl)^{(3-p) (7-p)\over 5-p}
 z^{(3-p) (7-p)\over 5-p}
~,
\eea
where $z=u+{1\over v}$.
{}From this it is straightforward to see that the near singularity background,
according to the definition given in the previous section, is
\bea
d\bar s^2&=&2 x^{-{3-p\over5-p}} dudv+x^{{7-p\over5-p}}ds^2(\bR^p)+x^{-{3-p\over5-p}}
d\Omega^2_{8-p}~, ~~~~~~~~x=v
\cr
\bar F_{p+2}&=&
x^{{9-p\over 5-p}}~  du\wedge dv\wedge d{\rm vol}(\bR^{p})
\cr
e^{2\bar\phi}&=&  x^{-{(3-p) (7-p)\over 5-p}}
~.
\eea
Note that in the case of null singularities as above the sign of the $u,v$ component
of the metric does not have an invariant meaning because it can change with a
coordinate transformation $u\rightarrow -u$.
The  exponents of the power-law
singularities of the Dp-branes, $p\not=3$, $p\leq 4$, are
\bea
ds^2&:&{\rm p}=-{3-p\over5-p}~,~~~~~~{\rm w}={7-p\over5-p}~,~~
~~~~{\rm q}=-{3-p\over5-p}
\cr
F_{p+2}&:& \ff={9-p\over 5-p}~,~~~~~p\leq 4
\cr
\phi&:& {\rm d}=-{(3-p) (7-p)\over 5-p}~.
\eea
The D3-brane
solution, although it is included in the above analysis, it is non-singular at $x=0$.
Similarly, the metric of the NS5-brane \cite{NS5Brane} is not singular at $r=0$, though the dilaton
diverges at $r=0$, and so we shall not investigate
them further. So from the Dp-branes, it remains to investigate the D5-branes.
In this case the transformation to conformal coordinates is $r=e^w$. Writing the
background in conformal coordinates $(\uu, \vv)$, $w=\uu+\vv, t=\vv-\uu$, we have
\bea
ds^2&=& 4 e^{\uu+\vv} d\uu d\vv+ e^{-\uu-\vv} ds^2(\bR^5)+ e^{\uu+\vv}  d\Omega^2_{3}
\cr
F_7&=&-4 e^{2\uu+2\vv} d\uu\wedge d\vv\wedge d{\rm vol}(\bR^5)
\cr
e^{2\phi}&=&e^{2\uu+2\vv}~.
\eea
The equation for the singularity is
\be
C=r=e^{\uu+\vv}=0
\ee
which has solutions either at $\uu=-\infty$ or at $\vv=-\infty$. We consider the case
where the singularity is located at $\vv=-\infty$ and define new conformal coordinates
$v=e^\vv$ and $u=e^\uu$. (The other case whether the
singularity lies at $\uu=-\infty$ leads to a near singularity geometry
with the same exponents.) In the $(u,v)$ conformal coordinates 
the background can be rewritten as
\bea
ds^2&=& 4  du dv+  u v ds^2(\bR^5)+ u v  d\Omega^2_{3}
\cr
F_7&=&-4  u v du\wedge dv\wedge d{\rm vol}(\bR^5)
\cr
e^{2\phi}&=& u^2 v^2~
\eea
and the singularity is at $u v=0$. Setting $x=v$, the near singularity geometry
for the D5-brane
is
\bea
d\bar s^2&=& 2  du dv+  x ds^2(\bR^5)+ x d\Omega^2_{3}
\cr
\bar F_7&=& x du\wedge dv\wedge d{\rm vol}(\bR^5)
\cr
e^{2\bar\phi}&=& x^2~
\eea
The  exponents of the power-law
singularities of the D5-brane are
\bea
ds^2&:&{\rm p}=0~,~~~~~~{\rm w}=1~,~~~~~~{\rm q}=1
\cr
F&:& \ff=1
\cr
\phi&:& {\rm d}=2~.
\eea
This concludes the computation of the near singularity geometries of the Dp-branes.

\subsection{Fundamental string}

The  fundamental string background \cite{FString} is
\bea
ds^2&=& H^{-1} (-dt^2+ d\rho^2)+ dr^2+ r^2 d\Omega^2_7~,~~~~~~~H=1+{Q_F\over r^6}
\cr
F_3&=&d{\rm vol}(\bR^{1,1})~dH^{-1}
\cr
e^{2\phi}&=&H^{-1}~.
\eea
Setting the mass per unit length, $Q_F$, of the string to $Q_F=1$,
the geometry near $r=0$
 is
\bea
ds^2&= &r^6 (-dt^2+ d\rho^2)+ dr^2+ r^2 d\Omega^2_7
\cr
F_3&=&d{\rm vol}(\bR^{1,1})\wedge dr^{6}
\cr
e^{2\phi}&=& r^6
~.
\eea
Transforming the two-dimensional metric $-r^6 dt^2+ dr^2$
into the conformal gauge with
the transformation $r=w^{-{1\over2}}$, we find
\bea
ds^2&=&w^{-3} (-dt^2+{1\over 4} dw^2)+ w^{-3} d\rho^2+ w^{-1} d\Omega^2_7
\cr
F&=&-3w^{-4} d{\rm vol}(\bR^{1,1})\wedge dw
\cr
e^{2\phi}&=& w^{-3}
~.
\eea
Performing the coordinate transformations
$z={1\over 2} w$, $\uu={1\over2}(-t+z)$ and $\vv={1\over 2}(t+z)$, we rewrite   the
above metric  as
\bea
ds^2&=&{1\over4} [2 z^{-3} d\uu d\vv+{1\over2} z^{-3} d\rho^2+ 2 z^{-1} d\Omega^2_7]
\cr
F&=& {3\over4} z^{-4} d\uu\wedge d\vv\wedge d\rho
\cr
e^{2\phi}&=&{1\over 8} z^{-3}~.
\la{middl}
\eea
To locate the singularity at $r=0$ in the new coordinates,
we note that $r= (2z)^{-{1\over2}}$
and so it lies at  $z=+\infty$. In  the conformal
coordinates $\uu, \vv$, the singularity
at $z=+\infty$
is located at  either $\uu=+\infty$ or $\vv=+\infty$. The two cases are symmetric
and so without loss of generality we can take the
singularity to lie at $\vv=+\infty$. The other case whether the
singularity lies at $\uu=+\infty$ leads to a near singularity geometry
with the same exponents.
 As we have explained,
we perform the conformal transformation $\uu=u$ and $\vv=1/v$ and set $x=v$.
In $(u,v)$ coordinates, the equation (\ref{middl}) can be written as
\bea
ds^2&=&{1\over4} [-2 z^{-3} v^{-2}  du dv+{1\over2} z^{-3} d\rho^2
+ 2 z^{-1} d\Omega^2_7]
\cr
F_3&=& -{3\over4} z^{-4} v^{-2} du\wedge dv\wedge d\rho
\cr
e^{2\phi}&=&{1\over 8} z^{-3}~,
\la{middla}
\eea
where $z=u+{1\over v}$.
{}From this it is easy to see that the
 near singularity geometry is
\bea
d\bar s^2&=& 2 x dudv+ x^{3} d\rho^2+  xd\Omega^2_7
\cr
\bar F_3&=& x^2 du\wedge dv\wedge d\rho
\cr
e^{2\bar\phi}&=&x^3~.
\eea
Therefore, the  exponents of the fundamental string solution are
\bea
ds^2&:& {\rm p}=1~,~~~~~~{\rm w}=3~,~~~~~~~~{\rm q}=1
\cr
F_3&:& \ff=2
\cr
\phi&:& {\rm d}=3~.
\eea
In the M-theory both the membrane solution \cite{DS2brane} and the five-brane
\cite{Guven} solution
are non-singular at the position of the branes, $r=0$. Because of
this we shall not present
the  exponents for these cases.

\subsection{Intersecting branes}

There are several intersecting brane backgrounds \cite{IT1, IT2, IT3}.
Here we shall focus only
in a few examples and present the near singularity geometries.
 In what follows we shall compute
the near singularity geometries of the metrics. A class of
intersecting brane configurations
in string theory are those of a fundamental string orthogonally
 ending on a Dp-brane, $p<6$.
The metric of a supergravity solution which
represents such a (delocalized) intersection in the string frame is
\bea
ds^2&=&- H_D^{-{1\over2}} H^{-1}_F dt^2+ H_D^{{1\over2}} H^{-1}_F d\rho^2
+ H_D^{-{1\over2}} ds^2(\bR^p)
+H_D^{{1\over2}} (dr^2+r^2 d\Omega^2_{7-p})
\cr
&&~~~~~~~~~~~~~~~~~~~~H_D=1+{Q_D\over r^{6-p}}~,~~~~~
H_F=1+{Q_F\over r^{6-p}}~,
\eea
where $H_D, H_F$ are the harmonic functions of the Dp-brane and the fundamental string,
respectively.
Setting the mass per volume parameters of the branes $Q_D=Q_F=1$, the metric near
the singularity at $r=0$ is
\be
ds^2=- r^{{3\over2} (6-p)} dt^2+ r^{{6-p\over 2}} \Bigr(d\rho^2+ds^2(\bR^p))\Bigl)
+ r^{-{6-p\over 2}} dr^2+ r^{{p-2\over 2}} d\Omega^2_{7-p}~.
\ee
As in the case of branes, we perform a coordinate transformation
$r= w^{-{2\over 10-2p}}$ to set the two-dimensional metric spanned by $t,r$
in the conformal gauge. The above metric can then be rewritten as
\be
ds^2= w^{-{3(6-p)\over 10-2p}}\biggr( -dt^2+ {4\over (10-2p)^2} dw^2\biggl)
+ w^{-{6-p\over 10-2p}} \Bigr(d\rho^2+ds^2(\bR^p))\Bigl)+ w^{-{p-2\over 10-2p}}
 d\Omega^2_{7-p}~.
\ee
As in the case of Dp-branes, we define a new coordinate
$z={2\over 10-2p} w$, $p<6$, and
conformal coordinates $\uu={1\over 2} (t+z), \vv={1\over2} (-t+z)$.
It is clear that the singularity at $r=0$ for $p\leq 4$ is
located at either $\uu=+\infty$
or $\vv=+\infty$.  The two cases are symmetric so we take
 that the singularity  lies at $\vv=+\infty$.
As we have explained to locate the singularity at the origin,
we perform a conformal transformation $\uu=u$
and $\vv=1/v$ and set $x=v$. After all these coordinate changes the singularity is
located at $x=0$ and the near singularity geometry is
\be
d\bar s^2= -4 w^{-{3(6-p)\over 10-2p}} v^{-2} du dv
+ w^{-{6-p\over 10-2p}} \Bigr(d\rho^2+ds^2(\bR^p))\Bigl)
+ w^{-{p-2\over 10-2p}} d\Omega^2_{7-p}~,
\ee
where $w={10-2p\over 2} (u+{1\over v})$.
Using the definition of the near
singularity geometry, one can immediately
read the  exponents as
\be
{\rm p}=-{2-p\over 10-2p}~,
~~~~{\rm w}_1={\rm w}_2={6-p\over 10-2p}~,~~~~
{\rm q}={2-p\over 10-2p}~,~~~~ p<5
\ee
{}For $p=5$, the transformation to conformal coodinates is $r=e^w=e^{\uu+\vv}$.
The singularity is located at either $\uu=-\infty$ or $\vv=-\infty$ and the
two cases can be treated symmetrically. If the singularity is located at
$\vv=-\infty$, we define new conformal coordinates $v=e^\vv, u=e^\uu$.
Using these, one can easily read the exponents
as
\be
{\rm p}={1\over2}~,
~~~~{\rm w}_1={\rm w}_2={1\over2}~,~~~~
{\rm q}={3\over2}~.~~~~
\ee

As another example of intersecting brane configuration in M-theory,
one can take two orthogonally
intersecting membranes at a 0-brane. The metric is
\bea
ds^2&=& H_1^{{1\over3}} H_2^{{1\over3}} \biggr( -H_1^{-1} H_2^{-1}dt^2
+ H_1^{-1} ds^2(\bR^2)
+H_2^{-1} ds^2(\bR^2)+ dr^2+ r^2 d\Omega^2_{5}\biggl)
\cr
&&~~~~~~~~~~~~~~~~~~~H_1=1+{Q_1\over r^4}~,~~~~~~~~~~~H_2=1+{Q_2\over r^4}~,
\eea
where $H_1, H_2$ are the harmonic functions associated with the membranes.
Setting again $Q_1=Q_2=1$, the near singularity geometry is
\be
ds^2=- r^{16\over3} dt^2+ r^{4\over3} (ds^2(\bR^2)+ ds^2(\bR^2))+ r^{-{8\over3}} dr^2
+r^{-{2\over3}} d\Omega^2_5~.
\ee
Changing coordinates as $r=w^{-{1\over3}}$, the above metric can be rewritten as
\be
ds^2= w^{-{16\over 9}} (-dt^2+{1\over9} dw^2)+ w^{-{4\over9}} (ds^2(\bR^2)+ds^2(\bR^2))
+ w^{{2\over9}} d\Omega^2_5~.
\ee
In the conformal coordinates $\uu={1\over2}(t+z)$, $\vv={1\over2}(-t+z)$,
 $z={1\over3}w$,
the singularity at $r=0$ is either located at $\uu=+\infty$ or
$\vv=+\infty$. In the latter
case, we perform another conformal transformation $\uu=u$ and
 $\vv=1/v$ to locate the singularity
at the origin. Setting $x=v$, it is easy to show that the exponents are
\be
{\rm p}=-{2\over 9}~,~~~{\rm w}_1={\rm w}_2={4\over 9}~,~~~~
{\rm q}=-{2\over 9}~.
\ee

Another example is the magnetic dual configuration of two M5-branes
 intersecting at the 3-brane.
The metric is
\bea
ds^2&=& H_1^{{2\over3}} H_2^{{2\over3}} \biggr( H_1^{-1} H_2^{-1}ds^2(\bR^{1,3})
+ H_1^{-1} ds^2(\bR^2)
+H_2^{-1} ds^2(\bR^2)+ dr^2+ r^2 d\Omega^2_{2}\biggl)
\cr
&&~~~~~~~~~~~~~~~~~~~H_1=1+{Q_1\over r}~,~~~~~~~~~~~H_2=1+{Q_2\over r}~,
\eea
Setting again $Q_1=Q_2=1$, the near singularity geometry is
\be
ds^2= r^{2\over3} ds^2(\bR^{1,3})+ r^{-{1\over3}} (ds^2(\bR^2)+ ds^2(\bR^2))
+ r^{-{4\over3}} dr^2
+r^{{2\over3}} d\Omega^2_2~.
\ee
In this case the transformation to conformal coordinates is $r=z=e^{\uu+\vv}$.
The singularity at $r=0$ is either at $\uu=-\infty$ or $\vv=-\infty$. The two cases
can be treated symmetrically by using the conformal transformation
 $v=e^\vv$ and $u=e^\uu$
to write the equation for the singularity as $u v=0$. In the case
where the singularity
is located at $v=0$, set $x=v$ and the exponents are
\be
{\rm p}=-{1\over 3}~,~~~{\rm w}_1={\rm w}_2=-{1\over 3}~,~~~~
{\rm q}={2\over 3}~.
\ee
The other case gives the same exponents.

\section{Power-law Singularities, Strings and M-theory}

In string  theory and M-theory novel mechanisms have been proposed
to resolved spacetime singularities.
Examples of such  mechanisms are the resolution of orbifold
singularities using the twisted sectors  \cite{pol},
 the resolution of
conifold singularities using D-branes \cite{strom}
and the resolution of the singularities
of planar Dp-brane solutions by lifting them to M-theory \cite{town}.
 So far such mechanisms have not been extended
to the context of black hole and cosmological singularities. Although there
is no concrete proposal how to resolve such singularities, there are mechanisms within
string theory that may allow a resolution.  For example, the higher curvature
corrections of string theory and M-theory can resolve singularities; for such a
proposal see eg \cite{tseytb}. However
in very few simple backgrounds such corrections have been computed
because of the lack of understanding of $\alpha'$ corrections to all orders.
Therefore, it is useful
\begin{itemize}
 \item to be able
to identify and interpret  the nature of a singularity in a string background,
 ie whether the singularity
is due to a brane placed in the background or to another object that
has an interpretation within string theory and

\item to  have a criterion
to estimate the severity of a spacetime singularity and the likelihood that
such singularity can be resolved in string theory.
\end{itemize}

We propose that the near singularity geometries defined in the previous sections provide
a way to identify the local nature of a singularity in a generic string background.
One expects
that as one goes near the singularity, the leading contribution to the metric and fluxes
 will be due to the matter placed at the singularity and the rest of the space
will contribute to subleading terms\footnote{The argument below  applies only in this case.}.
Since the near singularity geometry is constructed
by the leading contribution, it is expected that one can identify the local
 nature of the singularity
by looking at the near singularity geometry. For example, if  the power-law singularity
of a
supergravity background is the same as that of a planar D-brane, then one can conclude that this
singularity  is due to a planar D-brane.
Clearly this argument can be extended to other branes
and other objects in string theory that are singular in the effective theory and
their  solutions are known. In this way, we can have an understanding
of the nature of singularities that occur in a particular background.

Having identified and interpreted  a singularity
in a generic string background,
one can appeal to various mechanisms in string theory
 and M-theory to attempt to resolved it.
For example, if a singularity is due to Dp-branes, one
can conclude that near the
singularity it is more appropriate to use gauge theory
to describe the theory and
if a singularity is due to a planar fundamental string, then one expects such
a singularity to exist because of the presence of a fundamental object in the theory.

Using the arguments above and the fact that string theory is solvable
in some Penrose limit plane waves, we can give an estimate about
the severity of certain singularities and the likelihood that  these can be resolved
within string theory.
One may expect that the Penrose limits of a background can be taken
in a regular way which means
that if string theory is well-defined
in a supergravity background\footnote{We shall discuss the effect
of higher curvature terms at the end of the section.}, then it will be well-defined at
all its Penrose limits.
Since
free strings behave well at mild and  marginal ($\omega^2<1/4$) plane wave singularities,
it is an indication that singularities for which all their Penrose limits
are of mild or marginal ($\omega^2<1/4$) type may be resolvable within string theory.
The above arguments are not a proof
that string theory is consistent in backgrounds with
Penrose limits that exhibit
 mild or marginal ($\omega^2<1/4$)
singularities. It is only an indication that it may be so. Other
consistency checks should also be considered like for example
strong string coupling effects, backreaction and instantons. We have mostly
focus on the singularities of the metric. The singularities of the other fields
should also be analyzed as well.

Let us  now investigate the case  of backgrounds with timelike and null
power-law singularities  associated with plane waves, via a Penrose limit,  which
have   marginal ($\omega^2>1/4$) or severe type singularities.
There are such supergravity backgrounds, for example
there are
plane wave solutions with such profiles. One can argue that string theory
cannot resolve such singularities. (We shall discuss the issue of higher curvature
corrections at the end of the section.) As we have mentioned free string propagation
is singular in  plane waves with singularities of marginal
($\omega^2>1/4$) or severe type.
 Assuming that
there are no objects in string theory that their timelike or
null  near singularity geometries
have Penrose limits of the severe type, there does not seem to be an
interpretation of such singularities in string theory and so there are no
intrinsic string mechanisms that they can be used to resolve the singularity.
It is known  for example that for all branes the associated plane waves
are of marginal type with $\omega^2\leq 1/4$ \cite{bfp, fuji, ryang} 
(see also appendix
A).
{}From these one can conclude that timelike and null
power-law singularities which admit
a Penrose limit for which the associated  plane wave has a singularity of
marginal ($\omega^2>1/4$) and severe type cannot be resolved in string theory.
It is likely that a similar result holds for backgrounds with spacelike
singularities.

It remains to investigate  whether higher curvature corrections, like
$\alpha'$ corrections, can resolve spacetime singularities. The plane wave
backgrounds do not receive $\alpha'$ corrections \cite{klimcik, horow1,duval, tseytc},
so one does not expect
a singularity to be resolved {\sl after} taking the Penrose limit. As we have
explained in appendix B, there are several ways to  adapt the Penrose
limit after $\alpha'$ corrections are taken into account.
It is clear though that it is not possible to deduce from the Penrose
limit of a singular supergravity solution whether the associated string background,
after all $\alpha'$-corrections are taken into account, is singular or not.
However it is curious that in the case that the corrected metric depends
analytically on $\alpha'$, the limit that preserves the homogeneity of
the $\alpha'$-corrected field equations (see appendix B)
gives the same singular plane wave as that
of the Penrose limit of the singular supergravity background. This suggests
that either in the analytic case
the singularity cannot be resolved with $\alpha'$
corrections\footnote{ There does not seem to exist an example in the literature
of a singular supergravity background for which all $\alpha'$ corrections
are known and the corrected background is smooth.} or the singular
plane waves are limits in the space of deformations of such string backgrounds
which preserve the field equations. These suggest that  consistency of string theory
in such background would require consistency at the plane wave limit.

\subsection*{Acknowledgements}

I would like to thank M. Blau, U. Gran and M. O'Loughlin
for critically reading earlier versions of this paper and for
their comments. I would also like to thank A.A. Tseytlin
for many useful discussions and correspondence.
I am grateful to T. Damour and H. Nicolai 
for explaining their work to me.

\newpage

\appendix

\section{Penrose limits for power-law singularities}

There are three ways to compute the Penrose limits \cite{penrose, gueven, bfhp2} of a spacetime. The first method
was originally proposed by Penrose and uses adapted coordinates. This  has
recently been improved after the observation that one of the adapted
coordinates is the Hamilton-Jacobi function for null geodesics \cite{patricot}. The second
method uses a covariant definition  of the Penrose limit which identifies
the plane wave profile with either a certain component of the Riemann tensor evaluated
on a parallel transported frame along the null geodesic or with the frequencies
of the geodesic deviation equation for null geodesics \cite{papad1, papad2}. The third method is a
combination of the two methods above. It utilizes
 the Hessian of the Hamilton-Jacobi function evaluated
on a parallel transported frame along the null geodesic.
In particular, one defines the matrix
\be
B_{ab}= E_a^\mu E_b^\nu \nabla_\mu\partial_\nu S~,~~~~~~
g^{\mu\nu} \partial_\mu S\partial_\nu S=0~,~~~~~\mu,\nu=0,\dots, D-1
\la{wprof}
\ee
where $\nabla$ is the Levi-Civita connection, $S$ is the Hamilton-Jacobi function and
$E_+, E_-, E_a$, $a,b=1,\dots, D-2$, is a
parallel transported pseudo-orthonormal coframe along the null geodesic.
Note that ${\rm tr} B=\nabla^\mu\partial_\mu S$. The wave profile
is  given by
\be
A_{ab}=\dot B_{ab}+ \sum_c B_{ac} B_{cb}~,
\ee
where the derivative is with respect to the affine parameter, $\uuu$, of the null geodesic.
The associated plane wave at the limit is
\be
ds^2=2d\uuu d{\rm v}+ A_{ab}(\uuu) z^a z^b+ (dz^a)^2~.
\ee
The description of all these methods is given in \cite{papad2}. Here we shall use
the first  and third methods to compute the Penrose limits
of the near singularity geometries described in section two.

\subsection{Spacelike and timelike singularities}

Motivated by the form of near singularity geometries of brane configurations
in section three, we shall focus on the Penrose limits of the metrics
\be
ds^2= -2 x^p du dv+ \sum _i x^{w_i} ds^2(\bR^{n_i})+ x^q d\Omega^2_m~,~~~~~x=ku+\ell v~,
\ee
where $k, \ell=\pm 1, 0$ and the dimension of spacetime is
$D=\sum_i n_i+m+2$.
We shall first consider the case where $k\ell\not=0$.
In this case, we define a new  coordinate $y=k u-\ell v$ and rewrite the metric as
\be
ds^2={1\over2} \eta x^p (-dx^2+dy^2)+ \sum _i x^{w_i} ds^2(\bR^{n_i})+ x^q d\Omega^2_m~,
~~~~\eta=k \ell~.
\la{met}
\ee
The singularity is spacelike (timelike) for  $\eta=1$ ($\eta=-1$).

Using the rotation invariance of the above metric, we can write  the Hamilton-Jacobi function
$S$ as
\be
S= X(x)+P y+J_i z^i+ L\theta~,
\ee
where $z^i=z^{i1}$ and $\{z^{ir}; r=1, \dots, n_i\}$ are the standard coordinates
in $\bR^{n_i}$, and  $P$, $J_i$ and $L$ are the conserved momenta associated with the translation
invariance along $y$ and $z^i$, and the rotational invariance along the angle $\theta$ of
the sphere, respectively. The function $X$ is determined by solving the Hamilton-Jacobi
equation for null geodesics giving
\be
\biggr({d\over dx} X(x)\biggl)^2=P^2+{1\over2}\eta J_i^2 x^{-\ww_i+\pp}+{1\over2}\eta L^2
x^{-\qq+\pp}~.
\ee
The non-trivial null geodesic equations are
\bea
({\dot x})^2&=& 4 x^{-2\pp} \biggr({d\over dx} X(x)\biggl)^2=
4P^2 x^{-2\pp}+2\eta J_i^2 x^{-\ww_i-\pp}+2\eta L^2 x^{-\qq-\pp}~,~~~~~\eta=\pm 1
\cr
{\dot y}&=&2\eta x^{-\pp} P
\cr
{\dot z}^i&=& x^{-\ww_i} J_i
\cr
{\dot \theta}&=& x^{-\qq} L~.
\eea

{}For $\eta\not=0$, there are two distinct classes of null geodesics
to consider. The first class are those null geodesics for which $J_i=0$. In this case
the geodesic equations are
\bea
({\dot x})^2&=& 4 x^{-2\pp} \biggr({d\over dx} X(x)\biggl)^2=
4P^2 x^{-2\pp}+2\eta L^2 x^{-\qq-\pp}~,~~~~~\eta=\pm 1
\cr
{\dot y}&=&2\eta x^{-\pp} P
\cr
{\dot z}^i&=& 0
\cr
{\dot \theta}&=& x^{-\qq} L~.
\eea
The analysis of the Penrose limits in this case resembles that   already
 done in \cite{papad2}. The other class of null geodesics is that 
 for which one or more of the
 conserved charges $J_i\not=0$.

\subsubsection{ The $J_i=0$ case}

The leading behaviour of the $x$ geodesic equation in this case is
\bea
1.~~~{\dot x}&=& 2\; x^{-\pp}\; \tilde P~,~~~~\pp\geq \qq~;~~ \tilde P=P~(\pp>\qq)~;~~~
\tilde P^2= P^2+{1\over 2}\eta L^2~~~ (\pp=\qq)
\cr
2.~~~{\dot x}&=& \sqrt{2}\; x^{-{\pp+\qq+2\over2}}   L~,~~~~\qq> \pp~,~~~~\eta=-1
\la{geoa}
\eea
where $\tilde P^2, \tilde L^2\geq 0$ and we have chosen the plus sign 
in the equation for $x$.
These equations can be solved as
\bea
&1.&~~~ x^{\pp+1}= 2(\pp+1)\tilde P\;\uuu \; ~,~~~~\pp+1>0
\cr
&2.&~~~ x^{{\pp+\qq+2\over2}}= {\pp+\qq+2\over\sqrt{2}}\;L \;\uuu~,~~~~~~\pp+\qq+2>0~,
\la{geob}
\eea
where the inequalities in the exponents arise from the requirement 
that the singularity at $x=0$
is reached in finite affine time $\uuu$.

Because of the symmetries that preserve both the metric and the choice
of null geodesic, we expect that $B$ and so $A$ are diagonal.
The wave profile can be easily be computed  from
\bea
B_{ri, js}&=& E_{ir}^\mu E_{js}^\nu \nabla_\mu\partial_\nu S
=\delta_{ij} \delta_{rs} \partial_\uuu\log g_{ir,js}^{1\over2}
=\delta_{ij} \delta_{rs} \partial_\uuu\log x^{\ww_i\over2}
\cr
B_{\hat \alpha \hat\beta}&=&
 E_{\hat \alpha}^\mu E_{\hat \beta}^\nu \nabla_\mu\partial_\nu S=
\delta_{\hat\alpha \hat\beta} \partial_\uuu\log g^{{1\over2}}_{\hat\alpha \hat\alpha}
=\delta_{\hat\alpha \hat\beta} \partial_\uuu\log (x^{{\qq\over 2}} \sin\theta)~,
~~~~~\hat\alpha, \hat\beta=2,3\dots, m
\cr
B_{11}&=& {1\over {\sqrt g}} \partial_\mu (g^{\mu\nu} {\sqrt g} \partial_\nu S)-
\sum_i n_i B_{i1,i1}-(m-1) B_{22}=  \partial_\uuu \log (\dot x x^p x^{{q\over2}})~,
\eea
where $\dot x={d\over d\uuu}x$.
Using the formulae for the metric and the expression of the wave profile in terms
of $B$, we get
\bea
A_{ri,js}
&=& \delta_{ij} \delta_{rs} {\partial^2_\uuu g_{ir,js}^{1\over2}\over g_{ir,js}^{1\over2}}
=\delta_{ij} \delta_{rs} {\partial^2_\uuu x^{\ww_i\over2}\over x^{\ww_i\over2}}
\cr
A_{\hat \alpha \hat\beta}&=&\delta_{\hat\alpha \hat\beta}
{\partial^2_\uuu g^{{1\over2}}_{\hat\alpha \hat\alpha}\over
g^{{1\over2}}_{\hat\alpha \hat\alpha}}=\delta_{\hat\alpha \hat\beta} {\partial^2_\uuu
(x^{{\qq\over 2}} \sin\theta)\over x^{{\qq\over 2}} \sin\theta}=\delta_{\hat\alpha \hat\beta}
\biggr({\partial^2_\uuu x^{{\qq\over2}}\over x^{{\qq\over2}}}-{L^2\over x^{2\qq}}\biggl)
\cr
A_{11}&=&{\partial_\uuu^2 (\dot x x^p x^{{q\over2}})\over \dot x x^p x^{{q\over2}}}~.
\eea
It is easy to see that a non-homogeneous plane wave
 occur  when the Penrose limit is taken along null geodesics which exhibit
behaviour 2, ie $x(\uuu)\sim \uuu^a$ for $a={2\over \pp+\qq+2}$ ($\qq>\pp$), and
$\qq>\pp+2$.

We shall now investigate the behaviour of the null geodesics near the singularity and that of
the associated frequencies. We begin with spacelike singularities, $\eta=1$. Writing
the solution for null geodesics as $x\sim \uuu^a$, the various cases that
we consider are described in the table below:

\begin{equation}
\begin{array}{|c|c|c|c|}\hline
\mathrm{Conditions \; on\;} (P,L) & \mathrm{Constraints \; on\;}
(p,q) & \mathrm{Behaviour}& \mathrm{a} \\ \hline
~~i.~~P\neq 0, L = 0 & p > -1 & 1& {1\over \pp+1}\\
~ii.~~P=0, L\neq 0 & p+q > -2 & 2& {2\over \pp+\qq+2} \\
iii.~~P\neq 0, L \neq 0 & p>q, p >-1 & 1 & {1\over \pp+1}\\
~iv.~~P\neq 0, L \neq 0 & p<q, p+q >-2 & 2 & {2\over \pp+\qq+2}\\
~~v.~~P\neq 0, L \neq 0 & p=q>-1 & 1=2& {1\over \pp+1}\\ \hline
\end{array}
\end{equation}

It turns out that the components $A_{ir,js}\sim \uuu^{-2}$ and $A_{11}\sim \uuu^{-2}$
but $A_{\hat\alpha\hat\beta}\sim \uuu^{-\gamma}$, $\gamma\geq 2$.
The frequency squares $\omega_i^2$ and $\omega_\alpha^2$, $\alpha=1, \dots m$, where
$A_{ir,js}=-\delta_{ij}\delta_{rs} \omega_i^2 \uuu^{-2}$, $A_{11}=-\omega_1^2 \uuu^{-2}$ and
$A_{\hat\alpha\hat\beta}=-\delta_{\hat\alpha\hat\beta} \omega^2_{\hat\alpha} \uuu^{-\gamma}$, are
as follows:

(i)~~$P\neq 0, L = 0$. The frequencies are
\bea
\omega_i^2&=& {\ww_i\over 2(\pp+1)} \Bigr(1-{\ww_i\over 2(\pp+1)}\Bigl)~,
\cr
\omega^2_{\alpha}&=&{\qq\over 2(\pp+1)} \Bigr(1-{\qq\over 2(\pp+1)}\Bigl) ~, ~~~\gamma=2~,~~~~
\alpha=1,\dots,m~.
\eea

(ii)~~$P= 0, L \not= 0$.  The frequencies are
\bea
\omega_i^2&=& {\ww_i\over (\pp+\qq+2)} \Bigr(1-{\ww_i\over (\pp+\qq+2)}\Bigl)~,
\cr
\omega_1^2&=& {\pp\over (\pp+\qq+2)} \Bigr(1-{\pp\over (\pp+\qq+2)}\Bigl)~,
\cr
\omega_{\hat\alpha}^2&=&\cases{&$\frac{\qq}{(\pp+\qq+2)}
\Bigr(1-\frac{\qq}{(\pp+\qq+2)}\Bigl)~,~~~\qq<\pp+2~,~~~~~\gamma=2$
\cr
&$\frac{1}{4}+\frac{1}{2\qq^2}~,~~~~~~~~~~~~~~~~~~~~~~~\qq=\pp+2~,~~~~~\gamma=2$
\cr
&$2^{\frac{2\qq}{\pp+\qq+2}} {L^{\frac{2\pp-2\qq+4}{\pp+\qq+2}}\over
(\pp+\qq+2)^{\frac{4\qq}{\pp+\qq+2}}}~,~~~~~\qq>\pp+2~,~~~~~~
\gamma=\frac{4\qq}{\pp+\qq+2}$~.}
\eea

(iii)~~ $P\neq 0, L \neq 0$, ($\pp>\qq$). The frequencies are as in the case (i)
but the case $\pp+1=\qq$ with frequencies $\omega_1^2=\omega^2_{\hat\alpha}={1\over4}$
 does not arise because
$\pp+1>\qq$.

(iv)~~ $P\neq 0, L \neq 0$, $\pp<\qq$. The frequencies $\omega_i^2$ and $\omega_\alpha^2$
are as in the case (ii) but in addition $\pp<\qq$.

(v)~~~~$P\neq 0, L \neq 0$, $\pp=\qq$. The frequencies in this case are as in case (i)
after setting $\pp=\qq$.

\vskip 0.8cm

Next let us turn to the case of timelike singularities, $\eta=-1$. The various
behaviours are summarized
in the table below:

\begin{equation}
\begin{array}{|c|c|c|c|}\hline
\mathrm{Conditions \; on\;} (P,L) & \mathrm{Constraints \; on\;} (p,q)
&\mathrm{Behaviour} & \mathrm{a}\\\hline
~~i.~~ P\neq 0, L = 0 & p > -1 & 1& {1\over \pp+1}\\
~ii.~~ P=0, L\neq 0 & & - \\
iii.~~ P\neq 0, L \neq 0 & p>q, p >-1 & 1 & {1\over \pp+1}\\
~iv.~~ P\neq 0, L \neq 0 & p<q & - \\
~~v.~\sqrt{2}|P|>|L|  & p=q>-1 & 1=2 & {1\over \pp+1}\\ \hline
\end{array}
\end{equation}

The frequencies are as follows:

(i)~~~ $P\neq 0, L = 0$. The frequencies  are  the same as those
in case (i) for $\eta=+1$.

(iii)~~ $P\neq 0, L \neq 0,  \pp>\qq$. The frequencies  are  the same
as those in case (iii) for $\eta=+1$.

(v)~$P\neq 0, L \neq 0, \sqrt{2}|P|>|L|,   \pp=\qq$. The frequencies  are the same
as those in case (v) for $\eta=+1$.

The cases (ii) and (iv) do not occur because the null geodesics do not enter the
singularity. We have verified that the substituting the exponents for the D6-brane,
we recover the wave profile which has been computed in \cite{bfp, fuji, ryang}.

\subsubsection{ The $J_i\not=0$ case}

The leading behaviour  of the null geodesics in terms of the
affine parameter $\uuu$ near the singularity $x=0$ depends on
the various of the  exponents and there are several cases to consider.
In particular the leading behaviour of the $x$ geodesic equation is
\bea
1.~~~{\dot x}&=& 2\; x^{-\pp}\; \tilde P~~~~~~~~~~~~~~~~~~~\pp\geq \ww_i, \qq~,~~~~
\tilde P^2>0
\cr
2.~~~{\dot x}&=& \sqrt{2}\; x^{-{\pp+\ww_i\over2}}  \tilde J_i~~~
~~~~~~~~~\ww_i\geq \pp, \qq~,~~
~~~\tilde J_i^2>0
\cr
3.~~~{\dot x}&=& \sqrt{2}\; x^{-{\pp+\qq\over2}}  \tilde L~~~~~~
~~~~~~~~\qq\geq \ww_i, \pp~,~~~
~~~\tilde L^2>0~.
\la{geoc}
\eea
In case (1) (i) $\tilde P=P$, if $\pp> \ww_i, \qq$ ($\eta=\mp1$),
(ii) $\tilde P=\sqrt {P^2+{1\over2} \eta J_i^2}$, if $\pp=\ww_i>\qq$, (iii)
$\tilde P=\sqrt {P^2+{1\over2} \eta L^2}$, if $\pp=\qq>\ww_i$ and (iv)
$\tilde P=\sqrt{P^2+{1\over2} \eta J_i^2+{1\over2} \eta L^2}$, if $\pp=\ww_i=\qq$.
In case (2), (i) $\tilde J_i=J_i$, if $\ww_i>\pp, \qq$ ($\eta=+1$), (ii) $\tilde J_i=
\sqrt{J^2_i+L^2}$, if $\ww_i=\qq>\pp$ ($\eta=+1$) and the rest
of the possibilities are as in case (1).
In case (3), (i) $\tilde L=L$, if $\qq>\ww_i, \pp$ ($\eta=+1$) and the rest of
the possibilities are as either in case (1) or case (2).
The rest of the geodesic equations remain the same. (We have chosen the plus sign
in the equation for $x$).

The geodesic equations (\ref{geoc}) can be solved to yield
\bea
&1.&~~~x^{\pp+1}=2(\pp+1) \tilde P \uuu~~~~~~~~~~~~~~~~~~~~~~~~~\pp>-1
\cr
&2.&~~~x^{{\pp+\ww_i\over2}+1}= \sqrt{2}\; \tilde J_i\;({\pp+\ww_i\over2}+1)\;
\uuu~~~~~~~~~~~\pp+\ww_i>-2
\cr
&3.&~~~x^{{\pp+\qq\over2}+1}= \sqrt{2}\;\tilde L\; ({\pp+\qq\over2}+1)\;
\uuu~~~~~~~~~~~~~~\pp+\qq>-2~.
\la{solge}
\eea
The inequality restrictions on the  exponents arise from the requirement that
the spacetime is geodesically incomplete at $x=0$, ie the singularity
at $x=0$ is reached in finite affine time $\uuu$.

In the $J_i\not=0$ case, it is not apparent that the wave profile $A$ is diagonal
as the choice of null geodesic breaks some of the translational symmetries
in the $\bR^{n_i}$ directions. Because of this, it is rather involved to
compute the wave profile with the method we have used for the $J_i=0$ case.
Instead, we shall proceed with adapted (Penrose) coordinates and the
 Hamilton-Jacobi function
to give the plane waves that arise in the Penrose limits in Rosen coordinates.
The method has been explained in detail
in \cite{papad2}.
{}For this first observe that the non-trivial geodesic equations
 can be solved formally as
\bea
y&=&Y(\uuu, x_0)+y_0~,~~~~~ Y(\uuu, x_0)= 2\eta \int^\uuu
 d\lambda~ x^{-\pp}(\lambda, x_0) P
\cr
R(x)&=&\uuu+R(x_0)~,~~~~~~~~R(x)={1\over2}\int dx ~ x^\pp~
({d\over dx} X(x))^{-1}
\cr
z^i&=&Z^i(\uuu, x_0)+ z_0^i~,~~~ Z^i(\uuu, x_0)=J_i
 \int^\uuu d\lambda~ x^{-\ww_i}(\lambda, x_0)
\cr
\theta&=&\Theta(\uuu, x_0)+\theta_0~,~~~\Theta(\uuu, x_0)= L \int^\uuu
d\lambda~ x^{-\qq}(\lambda, x_0)
\eea
where $y_0, x_0, z_0^i$ and $\theta_0$ are integration constants.
The adapted coordinates are  $U=\uuu$,
\be
V=S(y_0,x_0, z_0^i, \theta_0)= P y_0+X(x_0)+ J_i z^i_0+L\theta_0
\ee
and the rest of the  integration constants  subject to  a gauge fixing condition.
It is convenient to chose as the  gauge fixing
condition
\be
P y_0+J_i z^i_0+L\theta_0=0~.
\ee
After taking the Penrose limit and doing some  computation, we find that the plane wave
in Rosen coordinates is
\bea
ds^2&=&2 dU dV-\eta{1\over2 P^2}x^\pp(U) ( J_i dz_0^i+ Ld\theta_0)^2
+ \sum_{i} x^{\ww_i}(U) ds^2(\bR^{n_i})
\cr
&+ &x^\qq(U) d\theta_0^2+ x^\qq(U) \sin^2(\theta(U))
\sum_{\alpha>1}^m d\phi_0^\alpha~,
\la{rose}
\eea
where $\phi_0^\alpha$ are the integration constants of the null geodesic
equations associated with the rest of the angular coordinates  that parameterize the m-sphere.
We have assumed that $P\not=0$. If $P=0$ and $L\not=0$, we can use the same gauge fixing condition
but now we solve it for $\theta_0$. If both $P=L=0$, we solve the gauge fixing condition
with respect to one of the $z^i_0$ for which the associated momentum $J_i\not=0$.
The plane wave metrics  in the various cases above can be easily computed and we shall
not present them here.

It is clear that the metric in Rosen coordinates (\ref{rose}) is off-diagonal and
so the transformation
to Brinkmann coordinates can be rather involved. This transformation is equivalent
to the construction of the parallel frame along the null geodesic. Because of the
non-diagonal nature of the metric, one expects that the solution for
the parallel transported  frame will in general involve
path-ordered-exponentials.
If the Rosen coordinates metric admits an additional
isometry, then the associated plane wave is a homogeneous space.
It may be that some of Penrose limits that arise are singular
homogeneous plane waves with non-vanishing rotation as those
of \cite{mmol}.


\subsection{Null singularities}

In the null case, we have  $k\ell=0$. If $k\not=0$, the metric, after a possible change
of coordinates, can be written as
\be
ds^2=2 x^\pp dudv+\sum_i x^{w_i} ds^2(\bR^{n_i})+ x^\qq d\Omega^2_m~, ~~~~~~x=k u~.
\ee
This  can be rewritten in terms of the $x$ coordinate as
\be
ds^2=2 x^\pp dx dv+\sum_i x^{w_i} ds^2(\bR^{n_i})+ x^\qq d\Omega^2_m~, ~~~~~~~
\ee
after replacing  $v$ with $kv$.
The case where $k=0$ and $\ell\not=0$ is symmetric and it will not be further
explained. The Hamilton-Jacobi function is
\be
S=-P^2 v+ J_i z^i+ L\theta + X(x)~,
\ee
where
\be
{d\over dx}T=\Bigr({J_i\over P}\Bigl)^2 x^{\pp-\ww_i}+\Bigr({L\over P}\Bigl)^2 x^{\pp-\qq}~.
\ee
The equations for the null geodesics are
\bea
{\dot v}&=&x^{-\pp} {d\over du}T
=\Bigr({J_i\over P}\Bigl)^2 x^{-\ww_i}+\Bigr({L\over P}\Bigl)^2 x^{-\qq}
\cr
{\dot z^i}&=& x^{-\ww_i} J_i
\cr
{\dot \theta}&=& x^{-\qq} L
\cr
{\dot x}&=&- x^{-\pp} P^2~.
\eea
The null geodesics reach the singularity $x=0$ at finite affine time $\uuu$
provided that $\pp+1>0$.

The wave profile $A$ is diagonal in this case. This can be seen by using
adapted coordinates as in the case with $J_i\not=0$ described in the previous
section. In particular, there is always a gauge fixing condition such that
the metric in Rosen coordinates is diagonal.
It turns out that
\bea
B_{ir,js}&=&\delta_{ij} \delta_{rs} \partial_\uuu \log g_{i1,i1}^{{1\over2}}
\cr
B_{\alpha \beta}&=&\delta_{\alpha\beta} \partial_\uuu
\log g_{\alpha\alpha}^{{1\over2}}~,~~~~~\alpha,\beta=1,\dots, m~.
\eea
The wave profile is
\bea
A_{ir,js}&=&\delta_{ij} \delta_{rs}
{\partial_\uuu^2g_{i1,i1}^{{1\over2}}\over g_{i1,i1}^{{1\over2}}}
=\delta_{ij}\; \delta_{rs}
 {\partial_\uuu^2 x^{{\ww_i\over2}}\over x^{{\ww_i\over2}}}
\cr
A_{11}&=&
{\partial_\uuu^2g_{11}^{{1\over2}}\over g_{11}^{{1\over2}}}
={\partial_\uuu^2 x^{{\qq\over2}}\over x^{{\qq\over2}}}
\cr
A_{\hat \alpha \hat\beta}&=&\delta_{\hat\alpha\hat\beta}
{\partial_\uuu^2g_{\hat\alpha\hat\alpha}^{{1\over2}}\over g_{\hat\alpha\hat\alpha}^{{1\over2}}}=
\delta_{\hat\alpha\hat\beta}
\biggr({\partial_\uuu^2 x^{{\qq\over2}}\over x^{{\qq\over2}}}-{L^2\over x^{2\qq}}\biggl)
~~~~~~~~\hat\alpha,\hat\beta=2,\dots,m
\eea
The components $A_{ir,js}$ and $A_{11}$ of the  wave profile behave
as  $A_{ir,js}\sim \uuu^{-2}$ and  $A_{11}\sim \uuu^{-2}$
but $A_{\hat\alpha\hat\beta}\sim \uuu^{-\gamma}$, $\gamma\geq2$.
The frequency squares $\omega_i^2$ and $\omega_\alpha^2$,
where $A_{ir,js}=-\omega_i^2 \delta_{ij} \delta_{rs} \uuu^{-2}$, $A_{11}=-\omega^2_1 \uuu^{-2}$ and
$A_{\hat\alpha\hat\beta}=-\omega_{\hat\alpha}^2
\delta_{\hat\alpha\hat\beta} \uuu^{-\gamma}$, are as follows:
\bea
\omega^2_i&=& {\ww_i\over 2(\pp+1)}(1-{\ww_i\over 2(\pp+1)})~,~~~
\cr
\omega_1^2&=& {\qq\over 2(\pp+1)}(1-{\qq\over 2(\pp+1)})~,~~~
\cr
\omega_{\hat\alpha}^2&=&\cases{ &${\qq\over 2(\pp+1)}(1-{\qq\over 2(\pp+1)})~,~~~~~
\qq\le\pp+1~~~~\gamma=2$
\cr
&${1\over 4}+{L^2\over (\pp+1)^2 P^4}~,~~~~~~~~~~~~\qq=\pp+1~,~~~~\gamma=2$
\cr
&${L^2\over ((\pp+1)P^2)^{{2\qq\over \pp+1}}}~,~~~~~~~~~~~~~\qq>\pp+1~,~~~~
\gamma={2\qq\over \pp+1}$~.}
\la{ffda}
\eea
In all the above case, we assume that $\pp>-1$ for the null geodesics
to reach the singularity $x=0$ at finite affine time.
Substituting the exponents of the Dp-branes, $p\leq 5$, into (\ref{ffda}),
we have verified that they reproduce the of the Dp-branes frequency squares which
have been computed using another method in \cite{bfp, fuji, ryang}. This is also the case for
 the frequency squares
of the  fundamental string solution which has originally  been  computed in \cite{bfp}.

\section{$\alpha'$-corrections and Penrose limits}

In string theory and M-theory, the effective supergravity theories are modified
by higher curvature terms.  To distinguish between the field equations
 before or after higher curvature corrections are included we shall refer to
 the former as supergravity field equations and to the latter as string or M-theoy field equations.
 String $\alpha'$-corrections are accompanied
  with an appropriate power of $\alpha'$ and so the metric, dilaton and
  the various form-field strengths
  of generic solutions of the string are expected
to depend on $\alpha'$.  For such backgrounds, there are three ways to take the Penrose limit:
\begin{enumerate}
\item{} One can take the Penrose limit of the original supergravity background
before the $\alpha'$ corrections are included.

\item{} One can take the usual Penrose limit of the string background after the
$\alpha'$ corrections are included.

\item{} One can take the usual Penrose limit which also involves a
rescaling of the $\alpha'$ as $\alpha'\rightarrow \Omega^2\alpha'$,
where $\Omega$ is a parameter and $\Omega^2\rightarrow 0$ at the limit \cite{bfp}.
\end{enumerate}

The limit in case (1) is consistent in the sense that the associated
plane wave will solve the string field equations. This is because
plane wave backgrounds do not receive $\alpha'$ corrections \cite{horow1, duval, tseytc}. The wave
profile is the matrix of frequency squares of null geodesic deviation equation of the
supergravity metric.

In case (2), the resulting plane wave will not necessarily solve the
string field equations. This is because the string field equations are not
homogeneous with respect to the $\Omega$-scaling necessary to take the Penrose limit.
The wave
profile is the matrix of frequency squares of null geodesic deviation equation of the
string metric and in general will be different from that of the associated
supergravity metric.

In case (3), the limit is a modification  of the Penrose limit. To take this modified limit,
one first puts the string metric in adapted (Penrose) coordinates and then performs the
usual rescaling of the coordinates with the parameter $\Omega$, as for the
Penrose limit. In addition one  rescales $\alpha'$ as described.
If the metric in adapted (Penrose) coordinates
depends analytically of $\alpha'$, the metric at the limit
is the one computed in (1) from the associated supergravity background.
The Penrose limit
commutes with the operation of $\alpha'$ corrections.
However, if the string metric
does {\ it not} depend analytically on $\alpha'$, then either the limit is ill defined
or it  gives a metric which is not always a plane wave. If the limit
is well-defined, the background at the limit will solve the string field
equations because they transform homogeneously under the scaling required
for (3).

To illustrate the features explained above consider the singular metric associated
with the $SL(2,\bR)_k/U(1)$ coset model \cite{dvv, sfetsos, tsemetr}
\bea
ds^2&=&{1\over2}(k-2) (-dt^2+ dr^2+\beta^2(r) d\theta^2)
\cr
{4\over \beta^2}&=&\tanh^2({r\over2})-{2\over k}
\la{str}
\eea
where the level $k\sim \alpha^{-1}$  and we have added a spectator time direction
for the metric to have  Lorentzian signature. This background has also non-vanishing dilaton but
for simplicity we shall not consider this here.
This metric is singular at $\tanh^2({r_0\over2})={2\over k}$.
The associated supergravity metric is found at the limit $k\rightarrow \infty$ as
\be
ds^2={k\over2} (-dt^2+ dr^2+4 \coth({r\over2}) d\theta^2)~.
\la{sgr}
\ee
The singularity of the supergravity metric is at $r=0$.
We shall examine the Penrose limits of the near singularity geometries
for the two  metrics above.

To take the Penrose limit as described in case (1), it suffices  to take the usual
Penrose limit for the supergravity metric (\ref{sgr}). It is easy to see that the near
singularity geometry in this case is
\be
ds^2={k\over2}(-dt^2+ dx^2+x^{-2} d\theta^2)~.
\ee
The plane wave profile has one independent component $A$ which can be easily
evaluated from the Laplacian of the Hamilton-Jacobi function of this background.
A short computation reveals that
\be
A=2 \uuu^{-2}~,
\ee
ie the frequency is $\omega^2=-2$.

To take the limit as described in case (2), it suffices to take the usual
Penrose limit for the metric (\ref{str}). It is easy to see that the near
singularity geometry in this case is
\be
ds^2={1\over2}(k-2) (-dt^2+ dx^2+x^{-1} d\theta^2)~.
\ee
A short computation reveals that
\be
A={3\over4} \uuu^{-2}~,
\ee
ie the frequency is $\omega^2=-{3\over4}$. Therefore in both cases, the resulting
plane waves are homogeneous but the frequencies are different.

Finally to take the limit as described in case (3), it suffices to observe that
the dependence of the metric (\ref{str}) in Penrose coordinates is analytic in $\alpha'$.
Therefore this limit gives the same plane wave as that computed in (i).

\newpage

\rnc{\Large}{\normalsize}

\end{document}